\documentclass[12pt,preprint]{aastex}

\def\gs{\mathrel{\raise0.35ex\hbox{$\scriptstyle >$}\kern-0.6em
\lower0.40ex\hbox{{$\scriptstyle \sim$}}}}
\def\ls{\mathrel{\raise0.35ex\hbox{$\scriptstyle <$}\kern-0.6em
\lower0.40ex\hbox{{$\scriptstyle \sim$}}}}

\shortauthors{Owen et al}
\shorttitle{Deep Survey of A2125 I.}

\begin{document}

\title{A Deep Radio Survey of Abell 2125 I.}
\title{Radio, Optical and Near-IR Observations}

\author{
Frazer\,N.\ Owen,\altaffilmark{1,2}, 
W.\,C.\ Keel,\altaffilmark{3,2},
M.\,J.\ Ledlow\altaffilmark{4,2,5} G.\,E.\ Morrison,\altaffilmark{6,2},
\& R. A. Windhorst\altaffilmark{7,2}
}
\altaffiltext{1}{National Radio Astronomy Observatory, P.\ O.\ Box O,
Socorro, NM 87801 USA.; fowen@nrao.edu The National Radio Astronomy
Observatory is facility of the National Science Foundation operated
under cooperative agreement by Associated Universities Inc.}
\altaffiltext{2}{Visiting astronomer, Kitt Peak National Observatory,
National Optical Astronomy Observatories, operated by AURA, Inc.,
under cooperative agreement with the National Science Foundation.}
\altaffiltext{3}{Dept. of Physics \& Astronomy, University of Alabama,
Tuscaloosa, AL 35487 USA}
\altaffiltext{4}{Gemini Observatory, Southern Operations Center, AURA,
Casilla 603, La Serena, Chile}
\altaffiltext{5}{Deceased 5 June 2004. We shall miss his cheerfulness,
unfailing good sense, and scientific industry.}
\altaffiltext{6}{National Optical Astronomy Observatories, 950 N.
Cherry Ave., Tucson, AZ 85719 USA}
\altaffiltext{7}{Arizona State University, Dept. of Physics \&
Astronomy, Tyler Mall PSF-470, Box 871504, Tempe, AZ 85287-1504 USA}

\setcounter{footnote}{7}

\begin{abstract}

	We present a description of deep radio, optical, and near IR 
observations taken with the VLA, the KPNO 2m and the KPNO 4m of the 
region containing the rich cluster of galaxies Abell 2125. The
reduction of each dataset is described. A catalog of radio sources
apparently not associated with members of Abell 2125 and the
associated R-band magnitudes is presented. 

\end{abstract}


\section{Introduction}

	Deep radio observations offer one of the 
important windows on the evolution of star formation
and black-hole-related activity as a function of
cosmological epoch. Radio, optical/NIR, FIR
and X-ray data taken together have the potential 
eventually to give us a well-constrained picture
of our history. Each window, by itself, has serious
weaknesses. The optical/NIR window is the best studied
and provides the most direct view. However, it suffers from
uncertainties due to dust obscuration, a fundamental 
part of star-formation. FIR observations give us a 
signal from the dust but leave uncertain the origin of
heating of the dust and fail to see into the heart of
the AGN or star-forming regions. In X-rays and radio
we can see behind the dust but we are measuring  secondary
emission which is not easy to relate physically to the
black hole physics or star-formation-rate (SFR). Thus we need
a combination of all this information to deduce the full
picture.

	Below $L_{20cm} \sim 10^{23}$ W Hz$^{-1}$, one can
show that locally the radio luminosity function is dominated
by synchrotron emission related to star-formation driven processes
 \citep[e.g.,][]{con92}. Above this luminosity most sources are
twin-jet, black-hole-driven sources. Below the  $10^{23}$ W Hz$^{-1}$
break point, the sources are mostly found in spirals rather than
ellipticals and the radio and FIR luminosities correlate very
well \citep[e.g.,][]{yun01}. Through the radio-FIR
correlation, one can estimate the SFR for the radio alone and can
use the ratio of radio to submillimeter emission to constrain the
redshift (assuming the empirical radio-FIR relation continues to hold
at any redshift) \citep{cy99}. Thus deep radio observations
with limiting flux densities $<<1$ mJy, offer a way to study star
formation in the distant universe.

	However, there are a lot of assumptions in the previous
paragraph. What one really needs is deep data in all four windows
to understand the picture. Where we have such information new puzzles
arise. For example, SCUBA submillimeter sources are usually
associated with faint radio sources. Using the superior radio
positions, one can make faint optical identifications and try to
measure redshifts. Often when an optical redshift can be obtained,
one finds an AGN spectrum \citep[e.g.,][]{led02}. Even so, the
redshift estimated from
the Carilli/Yun relation agrees with the measured value. Perhaps this
is a coincidence but it might suggest that bright black-hole-driven
AGN's are part of the star-formation process in galaxies. 

	We are interested both in the very distant star-formation/AGN
history as well as the same phenomena in intermediate-distance
rich clusters($0.2 \ls z \ls 0.4$). A deep radio survey in the
direction on such a cluster can be used for both purposes. This paper
reports deep radio observations in the direction of Abell 2125. Future
papers will report the combination of these data with observations in
other bands to study both problems.

	We began studying the Abell 2125 field with medium deep, 20cm, 
VLA C-array observations \citep{dow,o99}.
We compared Abell 2125 ( a richness class 4, blue cluster and $z \sim
0.25$) with Abell 2645 (an apparently similar cluster at the same
redshift but with  much redder galaxies). We found a much higher detection
rate of radio galaxies (27 vs 4) in Abell 2125. The detection rate in
A2645 is consistent with a normal AGN population in a lower redshift
cluster. The A2125 excess population occurs entirely with luminosities
below  $L_{20cm} \sim 10^{23}$ W Hz$^{-1}$ and thus seems consistent
with a star-forming population. However, only a small fraction of
the excess population have optical spectra consistent with enough
star-formation activity to explain the population entirely that way.
Either the star-forming activity is very well hidden in the majority
of the low radio luminosity sample or the some other explanation is
necessary for the radio activity.

	One possible clue is that Abell 2125 does not appear to
be a typical very rich cluster when examined in more detail in the
optical and X-ray bands. Optically the cluster consists of a central
concentration together with a extension to the southwest extending
at least 2 Mpc. X-ray observations \citep{wang}
show the same pattern and that the total X-ray luminosity is low
for such a rich cluster. Much of the excess X-ray population is
contained in the southwest extension, not the cluster core. Thus
perhaps these unusual properties have something to do with the radio
excess.

	To study the cluster further, we have made a much deeper radio
observations and higher resolution using all four VLA configurations.
The deep radio data has also motivated deep optical and NIR imaging,
optical spectroscopy \citep{mil04}, deep millimeter/submillimeter 
observations \citep{e03}, and a deep {\it Chandra} exposure
\citep{w03}.In some cases, these ancillary observations were not 
motivated by the cluster but by the possibility of studying the
background sample ({\it e.g.} the submillimeter survey).
 In this paper we document the techniques used in reducing
the radio and optical imaging data for this field. We also report
the radio data and R-band optical identifications for these data.
A related paper contains the results for the radio sources confirmed
to be associated with cluster members.

\section{Observations, Reduction and Cataloging}

	Observations were made with the VLA for 3.5 hours in the 1997
B configuration and for 27.5 hours in the 1998 A configuration.
Since the total
time is dominated by the A configuration, the final image
for analysis had a resolution about 1.5 arcsec.  The data
were all taken in spectra line mode 4, which provides seven
3.125 MHz channels in each of 2 IFs (centered at 1365 and
1435 MHz) and each of two polarizations. Five second
integration times were used in the A and B array. The integration
times and channel bandwidths were chosen to minimize 
tangential and radial smearing of the images away from the
field center. It would be better to use narrower channel
widths, and shorter integration times to eliminate such
effects totally but the mode we picked seems the best 
compromise given the current VLA correlator and data
handling system. Another important consideration for deep
VLA work at 20cm with the present system is that the system
temperature increases as function of zenith distance due
to increased ground pickup. For this reason almost all the
observations were scheduled within 3 hours of source transit
to minimize this effect.

\subsection{Calibration \& Editing}

	All radio reductions
use the AIPS package. Since the total
bandwidth is relatively large for a spectral line experiment,
for each daily observation we first split a bright point source 
calibrator from the raw database and applied a phase 
self-calibration. Then we used this otherwise, uncalibrated
source to calculate a bandpass correction before we proceeded
with the rest of the regular continuum calibration. Regular
continuum calibration then continued, using the bandpass
calibration to flatten the spectral response, and thus
avoid the effects of the raw, sloping spectral response.
Calibration was made in the standard way to the Baars flux 
density scale using 3C286 as the flux calibrator. The weights 
associated with each integration 
were also calibrated as a function of observed system temperature
and calibrated as part of the standard AIPS processing.
	The calibrated dataset was first clipped using the
AIPS program CLIP well above the total found flux density found
in the field from lower resolution observations in order to remove
a few high points before beginning the self-calibration process.

\subsection{Imaging \& Self-calibration}

	The entire primary beam was then imaged using the
AIPS task IMAGR and the 3D, multi-facet options. Besides
37 facets, each with $1000\times 1000$ points, to describe the primary
beam, another 37 facets, each with $500\times 500$ points, were centered
on all the remaining detectable sources outside of the central
region. This is particularly important at 20cm since the feed
is a compromise using a lens to illuminate the primary reflector.
This feed causes both a higher system temperatures as a function of
elevation and produces a high first sidelobe response for the
primary beam. The result of this is that many outlying sources
are detected in the first sidelobe, as well as a few further out
which need to be removed. This was accomplished using the 37
outlying facets. 
	After initial imaging, each detected source had a tight clean box
placed around it to limit cleaning to real features. This procedure
allows each image to cleaned down to the 1$\sigma$ level and the
resulting clean components to be used in the self calibration process. 
This process also eliminates almost all of the clean bias and does
not artificially reduce the noise on the image as happens when one
does not use boxes. 

	The images were self-calibrated, first just in phase,
and later in both amplitude and phase using the AIPS task
CALIB.  However, these images were
limited by the pointing changes on outlying bright sources. This
is primarily due to the fact that the two circular polarizations
have different pointing centers ({\it i.e.} beam squint) on the VLA
due to the slightly
off-axis location of the feeds. This combined with the alt-az
geometry of the telescopes causes effective gain on a source far
off-axis to vary with hour angle. Also the slightly different
frequencies of the two IF's causes a slightly different primary
beam size and thus a source far from the field center has an higher
gain at the lower frequency.
As a final step to correct approximately  for these two effects, the
uv datasets were split into their separate IF's (2), polarizations (2)
and HA ranges (4). Then they were imaged and self-calibrated
separately. Finally the eight images of each facet were combined,
weighting each by $1/rms^2$ to form an optimum final image. 
 
	The final images had an {\it rms} noise, before correction
for the primary beam, typically of 6.5 $\mu$Jy, somewhat larger
near very bright sources. The final full resolution images had
a Gaussian clean beam with a FWHM of $1.60\times 1.52$ arcseconds at a
position angle of 87.9 degrees. 

\subsection{Cataloging}

	The inner $25\times 25$ arcminute region of the final image
was chosen for cataloging and further study. Beyond this region the
losses due to the bandwidth smearing and the primary beam attenuation
were judged to be too large for useful analysis. 
	The AIPS program SAD was used for forming the initial
source lists down to peak flux densities of 20 $\mu$Jy/beam.  The
residual images from
SAD were then searched to find any remaining sources sources missed
by SAD. Lower resolution images with
$3\times 3$ arcseconds and $5\times 5$ arcseconds were also
constructed in order to search for lower surface brightness sources
and to check the the fitted angular sizes for possible larger
components than were detected on the full resolution images. 

	The full list of detected sources was then inspected and
fitted manually using JMFIT. The parameter BWSMEAR was used to
take into account the finite bandwidth of the individual 3.125 MHz
channels used for the observing. Without using thus parameter,
bandwidth smearing makes sources appear to be the 
(fractional bandwidth)$\times$(distance from the field center) in
size, elongated radially from the phase pointing center.  This results 
in typical apparent source sizes of a $\sim 2$
arcsec over the field analyzed and may account for some previous
studies with the VLA for finding that typical size for sources 
in their samples. Sources  which were found by JMFIT to
have zero as the minimum size of the major axis were assumed
to be unresolved. For such sources the maximum allowed size for
the source on either axis was adopted as the upper limit to the 
angular extent. For unresolved sources, simulations have
shown that the fitted peak flux density for the unconstrained,
Gaussian functions fitted with JMFIT is the best estimate to the total
flux density \citep{g02}. Thus for sources
with only upper limits to their size this estimate of the
total flux density is used. For significantly resolved
sources the total fitted flux density is adopted. Errors in the
flux density and position were calculated using the formalism of
\citet{con98}.

	The local noise was estimated locally using the
AIPS program, RMSD, in a $100\times 100$ region centered on each
pixel. This calculation was made on a mosaicked image of the entire
$25\times 25$ arcmin of the central field. The resulting noise image
was then corrected for the primary beam attenuation, the bandwidth
smearing and approximately for time averaging to produce a local
noise image for the survey. This result was then used to pick source
which had five sigma peaks in excess of the local noise on the full
resolution image. 

\subsection{Optical Imaging}

	Optical and NIR images for the field were obtained 
of the same field in June 2000, using MOSAIC on the KPNO 4m
and SQIID on the KPNO 2.1m. With MOSAIC, images were obtained
in 8 bands: U, B, V, R, I, BATC 8010A, BATC 9170A, and WR CIII. The
BATC filters \citep{f96,x02} are special medium bandwidth filters 
($\sim 250$\AA) 
designed to fit in the clearest optical windows in the red.
 WR CIII is a narrow-band filter centered
at 4253\AA\ with a FWHM of 52\AA; we used it to isolate [OII]
at the redshift of A2125. Total
exposure times were U: 6 hours, B: 2 hours, V: 2 hours, R: 4 hours,
I: 1 hour, 8010A: 2 hours, 9170A: 2 hours, WR CIII: 3.6 hours.  
The pointings were dithered to five separate pointing near the
field center and then flattened and combined using the standard
IRAF MOSAIC software as described by \citet{valdes}. The standards of
\citet{land}  were used to set the magnitude scales on the UBVRI images. 
 The USNO catalog of
objects from the Palomar Sky Survey was used to set the astrometry
on each final image. 
	Each final image has a pixel size of 0.26 arcsec and is
$8700 \times 8800$ pixels on a side. 

\subsection{Near IR Imaging}

	The SQIID images were taken simultaneously in J, H, and K.
Each image has a usable field of $~\sim 5.2\times 5.2$ arcmin.
Observations were obtained with SQIID over a 3x3 grid with 35'' overlap across 
adjacent fields yielding a mosaicked field of view of 16.5
arcminutes. The mosaic 
consisted of 24 co-adds of 10 sec at each grid pointing. We applied random 
offsets between successive iterations of the mosaic.  The total useful
exposure time on the field was about 7 hours.
More integration time was spend on the the central field and the field
to to east of the center than on the other seven pointings.

As mosaicking a large area of sky with SQIID is not a standard observational 
mode, we summarize our data reduction method here.  We first subtracted off a 
stacked dark exposure from all frames.  For the sky subtraction we created a 
list of exposures over a $\pm$30 minute window. Each frame was inspected and
those with bright stars or higher than average background were removed.  The 
dark-subtracted sky frame was then subtracted from all object frames in that 
time window. A master flatfield was created from all suitable exposures over the 
night and combined with sigma clipping and mode scaling.  After
flatfielding, we 
next removed any large-scale residual sky gradients by subtracting a 
median-filtered version of each frame. Next, each frame was corrected for 
geometrical distortion. The distortion map was determined separately for each 
filter based on a cross-correlation of objects with our deep R-band frame which 
had been previously re-gridded to a proper world coordinate system (distortions 
removed). Finally, we applied a bad-pixel mask to each frame which set bad 
regions to a very large positive number (100000) to facilitate threshold 
clipping during co-addition. Registering and stacking the data was accomplished 
by creating a zero-level image larger than the full field of the 3x3 grid (1800 
pixels) and pasting each exposure into this image template. We used our R-band 
image (same image scale) as the reference image for determining pixel shifts. 
After measuring the shifts for each of the 9 frames in the grid, the frames were 
shifted and co-added with inverse-variance weighting and an upper threshold value 
set to clip off the bad regions set by the masking.

\section{Results}

	Observations were taken of this field to study
both the cluster and noncluster radio populations in the field
of A2125.  The cluster
members were isolated using spectroscopic redshifts as reported
in \citet{mil04} and discussed in detail in \citet{o04b}. The 
remaining radio-detected objects are believed to
predominately background sources. First, most have optical 
identifications much too faint to be cluster members and/or
have spectroscopic redshifts outside the cluster limits. Second,
as reported in \citet{o04d}, our photometric redshift estimates
or limits for very faint objects based on the implied absolute
magnitudes for almost all the remaining objects are consistent with
much higher redshifts than A2125 ($z=0.247$). These sources and their
parameters we list below. We will supply {\it fits} images of the
radio and optical images on request.

\subsection{Radio Catalog}

	In table 1, we give the final source list and parameters
from the fits. Column (1) contains the source name. The first two
digits are the facet number and the last three a sequence number.
Columns (2) and (3) contain the radio RA and Dec along with the
estimated error. In column (4) we give the distance from the pointing
center (15 41 14.00, 66 15 00.0) in arcmin. Column (5) contains the
observed (uncorrected) peak flux density from the map in $\mu$Jy per
beam. In column (6) we list the corrected total flux density and
estimated error.
   In column (7), we give the best fit size in arcseconds. If a two dimensional
Gaussian was the best fit, we give the major and minor axis size
(FWHM) and the position angle. Upper limits are given for sources
which were unresolved. Sources with sizes only were estimated directly
from the maps using the AIPS task, TVSTAT. Column (8) contains the
R-band optical magnitude in an aperture of radius 2 arcsec. Finally, 
column (9) gives the optical/NIR identification status. ``i'' indicates an
identification
in at least one of the 10 observed bands. ``b'' indicates a blank
field. ``c'' indicates an optical field which is too confused to be
sure of the identification. For the full sample of 357 radio sources
82\% had a optical identification; 10\% were confused; and 8\%
were blank and not confused.

\subsection{Optical Identifications}

\begin{figure}
\plotone{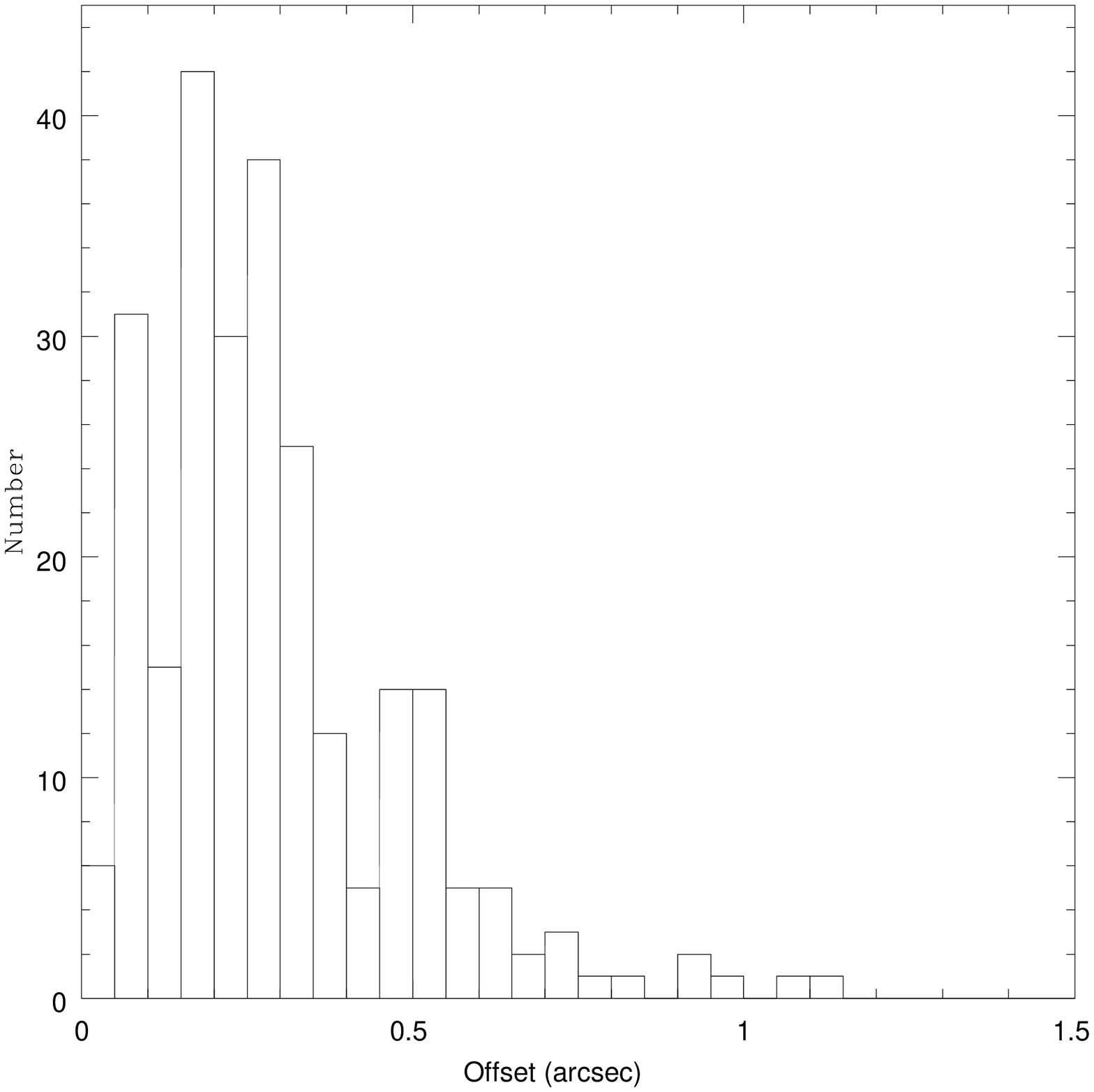}
\caption{Histogram of the offsets between the
positions of the radio sources and their optical identifications in arcsec.
\label{offsets}}
\end{figure}

	Most of the sources in the radio catalog have clear
identifications on the R-band image. For sources with apparent
IDs within 1 arcsecond of the radio position, the mean radio-optical
offset is -0.035 seconds of RA (-0.2 arcsec)  and -0.1 arcsec in
declination. 
After correcting for these offsets, the mean radio-optical separation is
0.3 arcsec. In 
figure \ref{offsets} we show the histogram of the total radio-optical offset
in arcseconds. Clearly most of the sources agree within a few tenths
of an arcsec. However, both some of the radio and some of the optical
sources are complex. Thus it seems prudent to catalog IDs out to
offsets of at least one arcsec. In the few cases where sources appear
much more extended than the beam, an attempt has been been made to
assign an obvious ID. However, it is difficult to assess the reality
of these IDs, statistically. 

\begin{figure}
\plotone{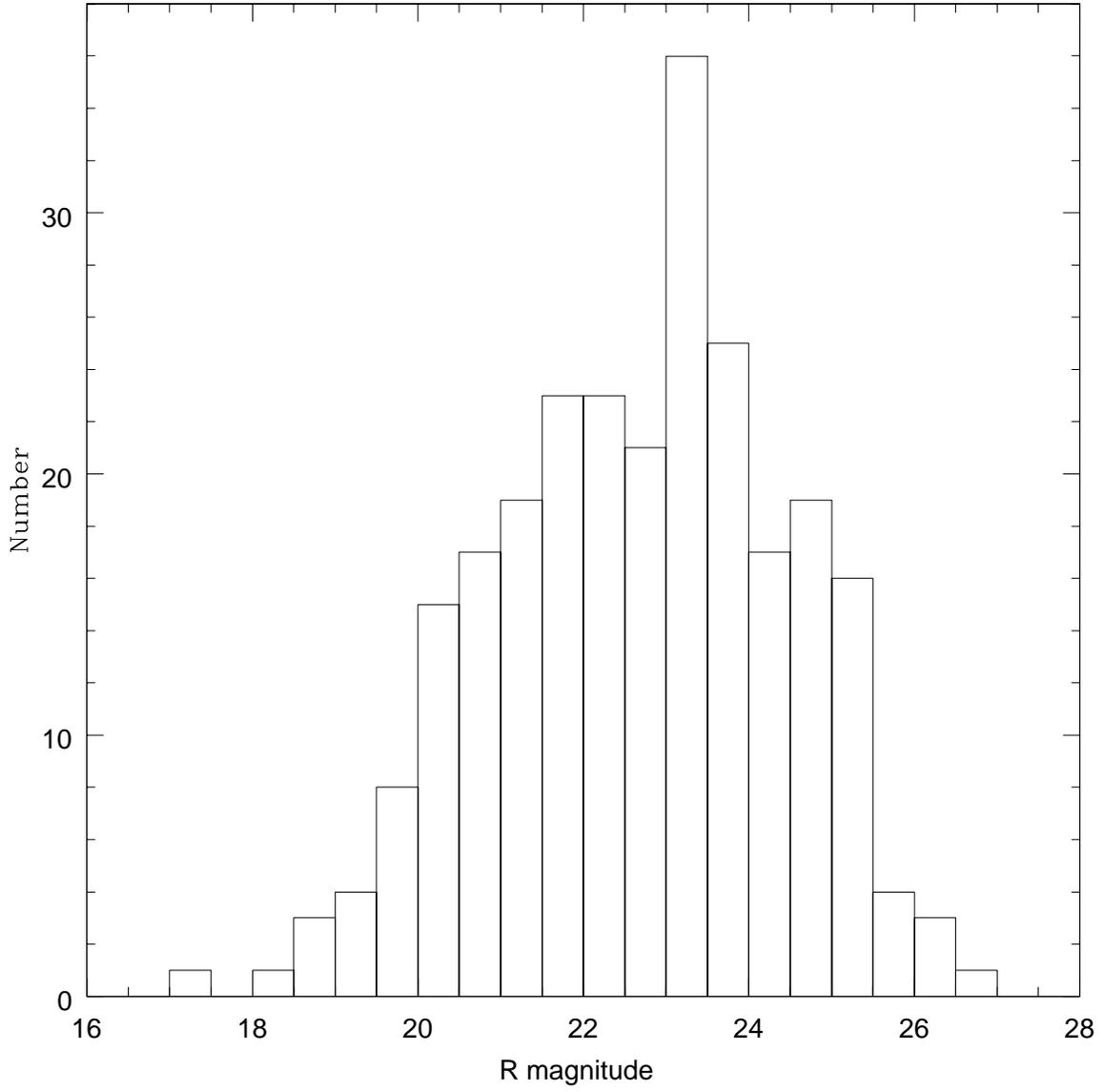}
\caption{Histogram of R magnitudes of identifications in an aperture
of two arcsec radius \label{rhist}}
\end{figure}

	In figure \ref{rhist} we show the histogram of optical magnitudes
at R (within a two arcsecond radius aperture) for
the identifications. The mean magnitude for the 257 objects with
identifications in the R-band image is 22.6. The median magnitude is
about 23.0 with some uncertainty due to the sources confused by very
bright objects and objects not detected in R due to image imperfections.

	In order to estimate the reliability of the IDs, SExtractor
was run on the central $8\times 8$ arcmin field of our R image, using
an aperture of two arcsec radius. Since this field contains the
central concentration of the cluster, the local source density should be
somewhat higher than  in the outlying fields. At a limiting
magnitude of $R=27.0$ (where we begin to be incomplete), we find 0.01
objects per square arcsec between 20.0 and 27.0. Down with a brighter
limit of $R=25.0$, we find 0.0016 objects per square degree. Given our 329
unconfused objects, we expect 10 false identifications down to 27.0.
However, 77\% of the objects have identifications brighter than 25.0.
Since any objects fainter than 25.0 would be too faint to be designated
the identification in the presence of a much brighter ID, we expect no
more than one of these to be incorrect. Between 25.0 and 27.0 we might 
expect one false ID. Neither of these
arguments take into account the full effect of the brighter
identifications and thus this estimate is probably pessimistic. The
argument does say that we expect a small number of cases where the
correct identification will have a fainter confusing source within
the one arcsec search area (and the seeing disk) as well.    

\section{Conclusion}

	We have presented the data reduction techniques and the
catalog of radio detections of non-cluster members
for the A2125 field. We also report the R-band optical
identifications. We report the
catalog of the detections for confirmed cluster members in a 
companion paper, as well as a more detailed interpretation of
these results. Subsequent papers will report more detailed
astrophysical results for this dataset.

\clearpage
\clearpage

\clearpage

\begin{deluxetable}{lrrrrrlrc}
\tablecolumns{9}
\tablewidth{0pt}
\tablecaption{Non-cluster Member Radio Sources\label{SL}}
\tablenum{1}
\pagestyle{empty}
\tablehead{
\colhead{Name} & 
\colhead{RA(2000.0)} &
\colhead{Dec(2000.0)} &
\colhead{Dist} &
\colhead{Peak} &
\colhead{Total} &
\colhead{Size} &
\colhead{Rmag} &
\colhead{ID}\\
\colhead{} &
\colhead{} &
\colhead{} &
\colhead{Mpc} &
\colhead{$\mu$Jy/b} &
\colhead{$\mu$Jy} &
\colhead{arcsec} &
\colhead{} &
\colhead{}}  
\startdata
18002&15 39 10.81(0.05)& 66 14 58.3(0.3)& 12.4&   69.9&  157.6(  16.)&$<$1.7          &22.44&i\\
11001&15 39 11.21(0.06)& 66 19 45.9(0.4)& 13.2&   70.0&  193.9(  28.)&   1.6x0p=118   & 0.00&b\\
11002&15 39 12.88(0.07)& 66 25 10.1(0.4)& 15.8&   38.8&  131.2(  24.)&$<$1.2          &23.48&i\\
18003&15 39 13.42(0.07)& 66 12 15.7(0.4)& 12.5&   63.9&  179.6(  28.)&   1.6x0.2p=71  &24.42&i\\
11003&15 39 14.08(0.08)& 66 22 20.2(0.5)& 14.1&   46.8&  169.4(  37.)&   1.9x0.3p108  & 0.00&b\\
18004&15 39 15.80(0.07)& 66 13 44.0(0.4)& 12.0&   57.5&  175.7(  31.)&   2.0x0.8p=55  & 0.00&c\\
24001&15 39 16.18(0.04)& 66 06 30.8(0.3)& 14.6&  136.2&  420.7(  27.)&   0.7x0p=50    &23.67&i\\
18005&15 39 16.24(0.07)& 66 12 33.9(0.4)& 12.1&   37.8&   77.9(  15.)&$<$1.8  & 0.00&c\\
24002&15 39 16.97(0.07)& 66 03 53.4(0.4)& 16.3&   38.1&  133.6(  25.)&$<$1.7          &24.39&i\\
18006&15 39 17.05(0.03)& 66 17 43.9(0.2)& 12.1&  267.0&  645.8(  29.)&   1.2x0p=103   & 0.00&c\\
11005&15 39 17.35(0.08)& 66 24 55.4(0.5)& 15.3&   43.7&  238.6(  60.)&   2.8x0.9p=138 &22.33&i\\
18007&15 39 17.45(0.07)& 66 11 15.4(0.4)& 12.3&   56.2&  170.8(  29.)&   1.7x0.5p=38  &22.30&i\\
11006&15 39 17.49(0.05)& 66 23 29.6(0.3)& 14.4&  110.9&  382.0(  34.)&   1.4x0p=167   &23.35&i\\
18008&15 39 17.62(0.03)& 66 18 50.3(0.2)& 12.3&  352.1&  854.3(  32.)&   0.9x0.1p=104 &22.48&i\\
11007&15 39 18.11(0.07)& 66 20 28.4(0.4)& 12.9&   55.8&  182.7(  32.)&   1.8x0.6p=125 &21.98&i\\
18009&15 39 18.62(0.07)& 66 16 59.7(0.4)& 11.8&   44.3&   89.0(  14.)&$<$1.9          &23.77&i\\
18010&15 39 19.69(0.08)& 66 17 16.3(0.5)& 11.7&   42.1&  110.7(  27.)&   1.8x0p=132   &23.00&i\\
18011&15 39 20.06(0.07)& 66 14 27.7(0.4)& 11.5&   43.2&   87.9(  14.)&$<$1.4          &20.67&i\\
18012&15 39 21.57(0.06)& 66 18 47.9(0.4)& 11.9&   48.8&  103.9(  15.)&$<$1.4          &25.30&i\\
18015&15 39 22.80(0.07)& 66 17 55.5(0.4)& 11.6&   42.8&   87.8(  15.)&$<$2.2          &22.40&i\\
18016&15 39 24.04(0.09)& 66 10 53.8(0.6)& 11.8&   35.4&   88.8(  28.)&   2.0x0p=90    &23.99&i\\
11008&15 39 25.19(0.07)& 66 26 28.1(0.4)& 15.8&   43.3&  145.1(  24.)&$<$1.5          &23.33&i\\
18017&15 39 26.08(0.07)& 66 11 29.3(0.4)& 11.4&   39.5&   76.6(  14.)&$<$1.3 &25.11&i\\
11009&15 39 26.16(0.08)& 66 20 54.1(0.5)& 12.3&   33.3&   73.9(  16.)&$<$2.0  &23.42&i\\
18019&15 39 31.14(0.04)& 66 17 01.9(0.2)& 10.5&  183.1&  337.6(  16.)&$<$0.7          & 0.00&c\\
18020&15 39 32.04(0.07)& 66 18 10.1(0.4)& 10.7&   40.1&   75.2(  13.)&$<$1.6          & 0.00&b\\
24004&15 39 33.09(0.05)& 66 07 43.6(0.3)& 12.5&  176.4&  817.8(  59.)&   4.5          &20.55&i\\
24006&15 39 35.03(0.03)& 66 08 03.8(0.2)& 12.2&  420.0& 1134.6(  51.)&   3.5          & 0.00&i\\
18021&15 39 35.56(0.03)& 66 14 56.7(0.2)&  9.9&  436.3& 1341.8(  61.)&  25.0          &23.72&i\\
18022&15 39 35.79(0.06)& 66 11 26.3(0.4)& 10.5&   48.9&   89.2(  13.)&$<$1.7          &19.80&i\\
11010&15 39 36.00(0.07)& 66 21 40.3(0.4)& 11.9&   35.7&   75.1(  15.)&$<$1.9          &21.70&i\\
11102&15 39 38.07(0.10)& 66 21 02.3(0.6)& 11.4&   33.6&  120.9(  47.)&   5.2x0p=48    &20.20&i\\
11011&15 39 39.05(0.07)& 66 23 13.4(0.4)& 12.6&   44.9&  101.4(  16.)&$<$1.2          &24.17&i\\
11012&15 39 39.46(0.04)& 66 22 45.2(0.2)& 12.2&  140.6&  306.0(  18.)&$<$1.2          & 0.00&i\\
11013&15 39 41.14(0.04)& 66 22 08.5(0.2)& 11.7&  153.2&  315.7(  17.)&$<$0.7          &24.94&i\\
18024&15 39 41.89(0.06)& 66 14 06.3(0.4)&  9.3&   58.3&   95.2(  12.)&$<$1.5  & 0.00&i\\
11014&15 39 42.55(0.08)& 66 24 55.8(0.5)& 13.5&   45.7&  169.8(  43.)&   2.7x0p=3     &19.53&i\\
18025&15 39 42.72(0.08)& 66 12 48.3(0.5)&  9.5&   33.5&   55.3(  12.)&$<$1.9  &23.96&i\\
18026&15 39 43.46(0.07)& 66 11 51.4(0.4)&  9.7&   39.4&   66.0(  12.)&$<$1.9          &25.16&i\\
11015&15 39 44.30(0.09)& 66 24 34.1(0.5)& 13.1&   47.4&  133.1(  35.)&   2.5x0p=135   &20.26&i\\
18027&15 39 45.42(0.02)& 66 12 36.5(0.1)&  9.2& 1403.2& 2312.5(  70.)&   0.5x0p=77    &21.18&i\\
11016&15 39 46.16(0.07)& 66 24 14.3(0.4)& 12.8&   39.4&   90.2(  16.)&$<$1.2          &21.98&i\\
24009&15 39 46.28(0.06)& 66 07 16.4(0.4)& 11.8&   55.2&  112.2(  15.)&$<$1.4          &21.55&i\\
24010&15 39 47.06(0.02)& 66 06 25.4(0.1)& 12.3&14041.0&82465.0(2475.)&  12.0          &21.65&i\\
24011&15 39 46.41(0.07)& 66 03 34.6(0.4)& 14.5&   61.5&  300.5(  51.)&   2.6x0.9p=61  &20.58&i\\
11017&15 39 46.43(0.08)& 66 21 25.4(0.5)& 10.9&   40.8&  114.1(  26.)&   1.8x0.7p=93  &21.52&i\\
18028&15 39 49.35(0.06)& 66 17 37.0(0.4)&  8.9&   58.3&   91.7(  11.)&$<$1.5          &20.60&i\\
18029&15 39 50.53(0.05)& 66 14 41.2(0.3)&  8.4&   97.4&  146.6(  11.)&$<$1.3          &23.37&i\\
11019&15 39 50.58(0.07)& 66 21 04.8(0.4)& 10.3&   41.3&   73.7(  13.)&$<$1.1          &22.38&i\\
24013&15 39 50.95(0.06)& 66 04 37.9(0.4)& 13.4&   54.0&  129.5(  17.)&$<$1.3          &26.45&i\\
24014&15 39 51.04(0.06)& 66 05 30.3(0.4)& 12.7&   74.3&  277.5(  40.)&   2.5          &22.49&i\\
11020&15 39 52.09(0.08)& 66 20 00.2(0.5)&  9.6&   45.4&  112.1(  25.)&   2.0x0p=57    &22.98&i\\
24039&15 39 52.24(0.07)& 66 07 15.6(0.4)& 11.3&   40.7&  468.3(  87.)&   6.5x3.3p=178 & 0.00&c\\
18031&15 39 53.22(0.09)& 66 18 32.8(0.5)&  8.9&   36.6&  173.0(  48.)&   5.6          &19.99&i\\
24017&15 39 53.25(0.04)& 66 10 43.0(0.3)&  9.2&  114.5&  182.7(  12.)&$<$0.9          &21.23&i\\
11021&15 39 53.82(0.06)& 66 20 44.4(0.4)&  9.9&   47.4&   80.8(  12.)&$<$1.7          &23.07&i\\
24018&15 39 54.54(0.02)& 66 05 12.9(0.1)& 12.7& 4864.9&27202.0( 818.)&   2.8x1.6p=100 &22.38&i\\
24019&15 39 55.74(0.07)& 66 08 38.0(0.4)& 10.2&   43.7&   75.5(  12.)&$<$1.3          &22.59&i\\
24020&15 39 56.01(0.08)& 66 05 32.7(0.5)& 12.3&   43.9&  190.2(  46.)&   3.1x1.0p=32  &19.17&i\\
11022&15 39 56.94(0.05)& 66 23 12.5(0.3)& 11.3&   94.5&  183.5(  15.)&$<$1.1          &23.76&i\\
11023&15 39 57.46(0.08)& 66 24 29.6(0.5)& 12.2&   34.7&   74.2(  15.)&$<$1.6 &21.31&i\\
11024&15 39 57.61(0.09)& 66 24 41.5(0.5)& 12.3&   34.5&  123.4(  36.)&   2.3x0.0p=25  &21.68&i\\
18032&15 39 57.87(0.03)& 66 13 31.6(0.2)&  7.8&  587.7& 1152.6(  46.)&   6.1          & 0.00&c\\
24023&15 39 57.97(0.05)& 66 07 49.4(0.3)& 10.5&   78.6&  140.3(  13.)&$<$1.5          &25.06&i\\
24024&15 39 58.10(0.05)& 66 03 17.4(0.3)& 14.0&   68.3&  176.4(  19.)&$<$1.4          &23.75&i\\
18034&15 39 59.73(0.06)& 66 17 04.0(0.3)&  7.7&   63.2&   89.8(  10.)&$<$1.3          &25.17&i\\
24025&15 40 00.01(0.02)& 66 05 51.6(0.1)& 11.8&12640.0&26159.0( 785.)&$<$0.3 &20.93&i\\
11025&15 40 01.07(0.08)& 66 22 16.7(0.5)& 10.3&   40.7&  104.9(  23.)&   1.6x0.8p=4   &23.23&i\\
11026&15 40 03.36(0.08)& 66 22 22.6(0.5)& 10.2&   44.0&  118.1(  29.)&   2.4x0p=172   &21.15&i\\
11027&15 40 04.21(0.07)& 66 25 23.4(0.4)& 12.5&   66.2&  201.9(  33.)&   2.2x0p=86    &22.05&i\\
18035&15 40 04.81(0.06)& 66 13 48.7(0.3)&  7.1&   80.1&  126.7(  14.)&   1.2x0p=163   &21.00&i\\
18036&15 40 04.89(0.06)& 66 13 02.9(0.4)&  7.2&   53.6&   72.9(  10.)&$<$0.9          & 0.00&b\\
24028&15 40 05.66(0.06)& 66 07 20.1(0.3)& 10.3&   61.6&  107.4(  13.)&$<$1.8          &23.75&i\\
24029&15 40 06.25(0.07)& 66 09 41.5(0.4)&  8.7&   50.9&  100.7(  18.)&   1.5x0.3p=59  &25.14&i\\
24031&15 40 06.94(0.06)& 66 02 50.6(0.4)& 13.9&   58.1&  148.0(  18.)&$<$1.7          &23.09&i\\
18037&15 40 07.27(0.09)& 66 15 59.9(0.5)&  6.8&   37.7&  104.4(  30.)&   3.4x0p=44    & 0.00&c\\
18038&15 40 07.79(0.09)& 66 17 13.4(0.6)&  7.0&   38.7&   95.7(  31.)&   4.7x0p=127   &21.80&i\\
18039&15 40 08.10(0.07)& 66 16 25.5(0.4)&  6.8&   54.5&   73.6(  12.)&   1.1x0p=33    &21.90&i\\
18040&15 40 08.70(0.05)& 66 15 36.2(0.3)&  6.6&   82.9&  107.8(  10.)&$<$0.7          &20.06&i\\
11028&15 40 10.44(0.06)& 66 24 38.5(0.4)& 11.5&   49.3&   97.7(  14.)&$<$2.0          &23.26&i\\
11029&15 40 10.82(0.09)& 66 19 22.3(0.5)&  7.7&   37.3&   92.2(  25.)&   2.5x0p=89    &21.46&i\\
11030&15 40 11.26(0.07)& 66 23 06.2(0.4)& 10.3&   38.1&   66.4(  12.)&$<$1.2          &21.57&i\\
24032&15 40 12.78(0.06)& 66 03 21.5(0.4)& 13.2&   48.7&  116.8(  17.)&$<$2.1          & 0.00&c\\
18044&15 40 13.88(0.07)& 66 18 29.4(0.4)&  7.0&   40.7&   54.2(   9.)&$<$1.2          & 0.00&c\\
18046&15 40 15.13(0.02)& 66 18 13.1(0.1)&  6.7& 1064.3& 1392.3(  43.)&$<$0.3          &23.11&i\\
18050&15 40 16.06(0.06)& 66 16 28.6(0.4)&  6.0&   58.8&   73.3(   9.)&$<$0.8          & 0.00&c\\
11033&15 40 21.63(0.04)& 66 24 15.1(0.2)& 10.6&  169.5&  304.5(  16.)&$<$1.0          &22.62&i\\
18051&15 40 21.82(0.08)& 66 17 54.9(0.5)&  6.0&   34.5&   44.0(   9.)&$<$1.1          & 0.00&i\\
18052&15 40 22.18(0.07)& 66 18 35.0(0.4)&  6.3&   38.5&   48.8(   9.)&$<$1.7      & 0.00&c\\
24034&15 40 22.73(0.06)& 66 06 30.3(0.4)& 10.0&   58.5&   97.5(  12.)&$<$1.7          &25.45&i\\
23035&15 40 22.93(0.06)& 66 10 01.7(0.4)&  7.2&   55.0&   73.2(  10.)&$<$1.1          &21.50&i\\
18053&15 40 23.44(0.07)& 66 14 16.1(0.4)&  5.1&   45.4&   53.4(   8.)&$<$0.9          & 0.00&c\\
23036&15 40 24.60(0.06)& 66 05 59.7(0.4)& 10.3&   50.9&   87.7(  12.)&$<$1.8          &21.51&i\\
11034&15 40 24.75(0.06)& 66 19 19.9(0.3)&  6.6&   64.4&   82.8(   9.)&$<$0.8          &22.68&i\\
18054&15 40 25.41(0.06)& 66 16 14.2(0.4)&  5.0&   52.9&   61.9(   8.)&$<$0.9          & 0.00&i\\
18055&15 40 26.53(0.08)& 66 15 30.3(0.5)&  4.8&   41.6&   60.7(  13.)&   1.4x0p=90    &20.81&i\\
24037&15 40 28.37(0.07)& 66 03 53.5(0.4)& 12.0&   43.2&   88.2(  15.)&$<$1.8          &20.95&i\\
11035&15 40 29.16(0.08)& 66 20 07.6(0.5)&  6.8&   34.0&   44.3(   9.)&$<$2.5  &21.63&i\\
11036&15 40 29.39(0.07)& 66 22 26.3(0.4)&  8.7&   75.7&  139.0(  21.)&   2.2x0p=66    & 0.00&b\\
11037&15 40 29.54(0.08)& 66 24 10.0(0.5)& 10.2&   33.2&   56.9(  12.)&$<$1.7  &21.24&i\\
18058&15 40 30.23(0.05)& 66 17 19.6(0.3)&  5.0&   76.3&   88.8(   9.)&$<$0.9          & 0.00&i\\
18059&15 40 30.46(0.07)& 66 15 30.9(0.4)&  4.4&   47.2&   58.2(  11.)&   1.2x0p=133   &23.77&i\\
18063&15 40 31.60(0.07)& 66 17 10.8(0.4)&  4.8&   39.8&   45.8(   8.)&$<$1.6  &21.69&i\\
12200&15 40 33.55(0.07)& 66 22 31.1(0.4)&  8.5&   33.0&   56.4(  11.)&$<$1.4  &23.26&i\\
25002&15 40 33.60(0.03)& 66 08 01.3(0.2)&  8.1&  516.0&  860.9(  29.)&   0.8x0.5p=12  &21.20&i\\
25003&15 40 33.80(0.05)& 66 07 24.3(0.3)&  8.6&   86.1&  127.3(  11.)&$<$1.3          &20.42&i\\
12001&15 40 33.91(0.07)& 66 20 03.8(0.4)&  6.5&   46.5&   59.1(   9.)&$<$1.6          &25.18&i\\
00003&15 40 35.24(0.05)& 66 12 04.6(0.3)&  4.9&   84.8&   97.6(   9.)&$<$1.0          &21.84&i\\
00004&15 40 35.56(0.03)& 66 16 56.6(0.2)&  4.3&  407.3&  457.6(  16.)&$<$0.5          &23.80&i\\
12002&15 40 36.18(0.06)& 66 20 46.9(0.4)&  6.9&   48.9&   64.0(   9.)&$<$1.0  & 0.00&b\\
00005&15 40 36.74(0.04)& 66 15 41.4(0.2)&  3.8&  181.9&  199.5(  10.)&$<$0.7          &22.93&i\\
25005&15 40 37.75(0.06)& 66 07 48.3(0.4)&  8.1&   57.4&   81.1(  10.)&$<$1.7          &21.16&i\\
25006&15 40 37.75(0.07)& 66 08 23.4(0.4)&  7.6&   69.3&  168.9(  30.)&   3.8x0p=154   &21.24&i\\
00006&15 40 37.85(0.07)& 66 12 32.7(0.4)&  4.4&   38.7&   43.3(   8.)&$<$0.9  & 0.00&b\\
12003&15 40 38.03(0.07)& 66 20 18.4(0.4)&  6.4&   35.7&   45.2(   9.)&$<$1.2  &22.06&i\\
25007&15 40 38.47(0.07)& 66 08 33.3(0.4)&  7.4&   43.8&   58.8(  10.)&$<$1.3          &23.46&i\\
25008&15 40 40.06(0.03)& 66 06 30.6(0.2)&  9.2&  517.7&  799.7(  26.)&$<$0.4          &23.61&i\\
12005&15 40 40.22(0.05)& 66 21 09.3(0.3)&  7.0&  101.1&  132.9(  10.)&$<$1.0          &24.50&i\\
12069&15 40 40.79(0.05)& 66 26 54.1(0.3)& 12.4&   79.8& 1848.0( 150.)&  20.0          &17.40&i\\
12006&15 40 41.24(0.05)& 66 25 25.6(0.3)& 10.9&   88.7&  162.3(  14.)&$<$1.1          &24.32&i\\
25009&15 40 42.55(0.03)& 66 08 54.1(0.2)&  6.9&  320.8&  414.8(  15.)&$<$0.5          &20.87&i\\
00009&15 40 44.24(0.07)& 66 16 54.2(0.4)&  3.5&   43.6&   55.4(  11.)&   1.2x0p=147   &21.86&i\\
25012&15 40 44.91(0.02)& 66 09 04.8(0.1)&  6.6&36362.0&46134.0(1384.)&$<$0.1          &22.09&i\\
12009&15 40 45.81(0.09)& 66 23 31.9(0.5)&  9.0&   37.1&   82.7(  25.)&   2.4x0p=74    &22.62&i\\
12010&15 40 46.43(0.07)& 66 21 01.6(0.4)&  6.6&   47.0&   59.9(   9.)&$<$1.2  &23.81&i\\
12011&15 40 46.44(0.05)& 66 26 19.4(0.3)& 11.7&   94.2&  185.1(  15.)&$<$1.1          &24.06&i\\
00011&15 40 47.19(0.05)& 66 15 51.8(0.3)&  2.8&  102.2&  107.7(   8.)&$<$1.0          &24.51&i\\
00012&15 40 47.32(0.07)& 66 11 37.9(0.4)&  4.3&   39.0&   43.4(   8.)&$<$1.2  &22.65&i\\
00013&15 40 47.61(0.09)& 66 12 08.7(0.5)&  3.9&   36.4&   43.1(  12.)&   1.6x0p=119   &24.82&i\\
12012&15 40 47.78(0.07)& 66 20 56.2(0.4)&  6.5&   45.1&   56.8(   9.)&$<$1.0          &23.14&i\\
12013&15 40 48.13(0.04)& 66 24 02.7(0.3)&  9.4&  121.2&  191.9(  12.)&$<$0.7          &21.95&i\\
12014&15 40 48.42(0.05)& 66 22 58.6(0.3)&  8.4&   68.6&   99.5(  11.)&$<$0.9          &22.79&i\\
12015&15 40 48.99(0.09)& 66 20 24.2(0.5)&  6.0&   36.2&   56.2(  15.)&   1.7x0p=0     &24.92&i\\
12016&15 40 49.87(0.07)& 66 20 15.1(0.4)&  5.8&   53.7&  200.3(  40.)&   8.0x0p=9     & 0.00&c\\
12017&15 40 49.57(0.06)& 66 21 05.5(0.4)&  6.6&   56.1&   71.2(   9.)&$<$1.2          &21.74&i\\
00016&15 40 50.13(0.06)& 66 16 40.2(0.4)&  2.9&   54.4&   75.4(  11.)&   1.1x0.5p=90  & 0.00&c\\
00015&15 40 50.15(0.04)& 66 10 45.7(0.2)&  4.9&  176.1&  236.1(  13.)&   1.0x0p=140   &21.89&i\\
00017&15 40 50.44(0.02)& 66 10 55.4(0.1)&  4.7& 4413.5&11876.0( 357.)&   2.8x1.1p=13  &22.17&i\\
00018&15 40 50.51(0.07)& 66 13 01.0(0.4)&  3.1&   40.7&   43.1(   8.)&$<$0.9  &21.77&i\\
00019&15 40 50.98(0.06)& 66 16 32.3(0.3)&  2.8&   84.4&   89.9(  10.)&   1.0x0p=7     &18.80&i\\
25014&15 40 51.79(0.03)& 66 06 30.9(0.2)&  8.8&  256.8&  383.0(  16.)&$<$0.6          &19.28&i\\
00021&15 40 52.60(0.06)& 66 16 36.6(0.4)&  2.7&   52.3&   54.7(   8.)&$<$1.1          &24.96&i\\
00023&15 40 53.00(0.09)& 66 16 54.4(0.5)&  2.8&   37.2&   40.2(  12.)&   1.6x0p=51    &23.35&i\\
00024&15 40 53.31(0.08)& 66 19 01.5(0.5)&  4.5&   46.8&   95.6(  19.)&   2.2          & 0.00&c\\
12020&15 40 54.24(0.05)& 66 21 31.7(0.3)&  6.8&   69.1&   89.1(   9.)&$<$1.4          &20.12&i\\
00025&15 40 54.43(0.07)& 66 12 50.3(0.4)&  2.9&   47.0&   49.4(   8.)&$<$1.0          &22.71&i\\
12021&15 40 54.68(0.08)& 66 20 48.5(0.5)&  6.1&   40.0&   74.8(  18.)&   1.9x0.1p=137 & 0.00&i\\
12022&15 40 54.91(0.07)& 66 20 43.8(0.4)&  6.0&   47.3&   57.9(   9.)&$<$1.7  &22.71&i\\
12023&15 40 55.29(0.03)& 66 21 52.0(0.2)&  7.1&  290.2&  426.5(  17.)&   0.7x0.4p=52  &22.77&i\\
12024&15 40 55.51(0.06)& 66 25 19.1(0.4)& 10.5&   47.7&   83.2(  12.)&$<$1.3          &24.15&i\\
12025&15 40 55.87(0.07)& 66 27 09.3(0.4)& 12.3&   42.8&   89.7(  15.)&$<$1.4          &23.38&i\\
00029&15 40 56.37(0.06)& 66 16 28.2(0.4)&  2.3&   52.0&   53.7(   7.)&$<$0.8          &22.36&i\\
12026&15 40 56.72(0.10)& 66 24 24.5(0.6)&  9.6&   36.0&  114.5(  40.)&   5.8x1.8p=7   & 0.00&i\\
12071&15 40 56.73(0.09)& 66 26 25.2(0.5)& 11.5&   36.4&   98.9(  30.)&   2.3x0p=36    &23.23&i\\
00030&15 40 56.88(0.05)& 66 17 26.5(0.3)&  3.0&   91.4&  128.6(  12.)&   1.1x0.6p=33  &23.15&i\\
25016&15 40 57.38(0.04)& 66 06 39.2(0.3)&  8.5&  122.2&  178.2(  12.)&$<$1.1          &24.52&i\\
12028&15 40 59.42(0.07)& 66 20 27.4(0.4)&  5.6&   51.3&   78.7(  15.)&   1.6x0p=71    &25.21&i\\
00032&15 40 59.59(0.07)& 66 12 16.4(0.4)&  3.1&   49.2&   83.6(  16.)&   1.9x0.2p=56  &22.10&i\\
00033&15 41 00.63(0.06)& 66 13 53.9(0.4)&  1.7&   85.8&  185.8(  28.)&   6.7          &21.29&i\\
00035&15 41 00.85(0.04)& 66 18 00.4(0.2)&  3.3&  133.3&  142.0(   9.)&$<$0.7          &25.18&i\\
12030&15 41 00.85(0.06)& 66 24 52.7(0.3)& 10.0&   62.8&  104.3(  12.)&$<$1.2          & 0.00&i\\
00037&15 41 01.06(0.05)& 66 14 16.3(0.3)&  1.5&   81.0&  113.3(  11.)&   1.1x0.8p=13  & 0.00&b\\
00038&15 41 01.52(0.07)& 66 13 54.2(0.4)&  1.7&   42.2&   42.9(   7.)&$<$1.3          &20.61&i\\
25018&15 41 02.03(0.02)& 66 07 11.5(0.1)&  7.9&  711.0&  986.8(  31.)&$<$0.3          &23.01&i\\
00040&15 41 02.19(0.07)& 66 18 15.0(0.4)&  3.5&   38.4&   41.1(   8.)&$<$1.2          & 0.00&i\\
00041&15 41 02.89(0.07)& 66 13 05.7(0.4)&  2.2&   44.2&   44.4(   7.)&$<$1.0          & 0.00&c\\
12031&15 41 03.07(0.07)& 66 26 31.1(0.4)& 11.6&   45.3&   87.8(  14.)&$<$1.8          &22.22&i\\
00042&15 41 04.65(0.05)& 66 14 56.4(0.3)&  0.9&   68.9&   69.3(   7.)&$<$1.2          &24.23&i\\
00043&15 41 05.35(0.06)& 66 12 35.6(0.4)&  2.6&   64.2&  124.9(  18.)&   2.1x0.8p=53  &19.27&i\\
12032&15 41 06.38(0.06)& 66 23 33.3(0.4)&  8.6&   55.7&   81.7(  11.)&$<$1.4          & 0.00&i\\
12034&15 41 07.31(0.04)& 66 21 23.7(0.2)&  6.4&  161.6&  202.0(  11.)&$<$1.2          &21.35&i\\
12035&15 41 07.45(0.06)& 66 23 35.5(0.4)&  8.6&   57.7&   84.6(  11.)&$<$1.6          & 0.00&c\\
00044&15 41 07.61(0.07)& 66 16 56.8(0.4)&  2.0&   44.9&   46.0(   7.)&$<$1.4          &20.84&i\\
00045&15 41 09.25(0.07)& 66 14 43.8(0.4)&  0.5&   42.7&   42.8(   7.)&$<$1.4 & 0.00&i\\
25020&15 41 09.39(0.06)& 66 08 07.3(0.4)&  6.9&   72.7&  105.1(  13.)&   1.1x0p=2     &23.80&i\\
12037&15 41 09.72(0.05)& 66 23 25.7(0.3)&  8.4&   81.1&  117.7(  11.)&$<$0.8          & 0.00&c\\
00048&15 41 09.91(0.07)& 66 10 58.7(0.4)&  4.0&   39.5&   43.2(   8.)&$<$1.7  &23.93&i\\
12038&15 41 10.04(0.06)& 66 22 10.5(0.4)&  7.2&   57.3&   75.5(   9.)&$<$1.3          & 0.00&c\\
12039&15 41 10.55(0.07)& 66 21 44.7(0.4)&  6.8&   38.5&   49.2(   9.)&$<$1.5  & 0.00&i\\
00049&15 41 10.63(0.05)& 66 13 13.9(0.3)&  1.8&  104.0&  105.9(   8.)&$<$0.9          &20.34&i\\
00050&15 41 10.75(0.07)& 66 12 09.7(0.4)&  2.9&   37.6&   39.4(   7.)&$<$1.5  &22.25&i\\
12040&15 41 10.95(0.07)& 66 27 07.8(0.4)& 12.1&   60.9&  126.1(  23.)&   1.7x0p=4     & 0.00&b\\
12041&15 41 11.08(0.07)& 66 24 35.8(0.4)&  9.6&   53.1&  129.2(  22.)&   1.7x0.4p=107 &24.54&i\\
00053&15 41 12.47(0.06)& 66 17 16.7(0.4)&  2.3&   55.9&   57.6(   7.)&$<$1.5          &21.67&i\\
00054&15 41 12.87(0.07)& 66 15 02.0(0.4)&  0.1&   37.0&   37.0(   7.)&$<$0.9  &21.71&i\\
00055&15 41 13.51(0.07)& 66 13 03.9(0.4)&  1.9&   41.6&   42.5(   7.)&$<$0.9          &23.48&i\\
25107&15 41 13.87(0.07)& 66 09 02.2(0.4)&  6.0&   37.7&   50.7(   9.)&$<$1.0  & 0.00&b\\
00058&15 41 14.78(0.04)& 66 13 30.7(0.2)&  1.5&  134.6&  136.4(   8.)&$<$0.7          & 0.00&i\\
12043&15 41 15.31(0.06)& 66 25 29.1(0.3)& 10.5&   63.3&  110.0(  13.)&$<$1.1          &23.09&i\\
12044&15 41 16.91(0.06)& 66 26 16.1(0.4)& 11.3&   60.8&  166.4(  23.)&   1.3x1.0p=8   &23.50&i\\
00061&15 41 16.97(0.06)& 66 16 26.7(0.3)&  1.5&   74.8&  188.1(  22.)&   2.5x1.2p=90  &20.80&i\\
12045&15 41 17.85(0.05)& 66 22 33.7(0.3)&  7.6&  104.2&  141.0(  10.)&$<$0.7          & 0.00&c\\
25023&15 41 20.72(0.07)& 66 09 33.5(0.4)&  5.5&   38.4&   46.2(   9.)&$<$1.8  &24.99&i\\
12046&15 41 20.80(0.08)& 66 19 20.7(0.5)&  4.4&   40.4&  140.3(  33.)&   4.3x2.0p=48  &19.88&i\\
00063&15 41 22.47(0.05)& 66 18 58.0(0.3)&  4.1&   72.5&   79.5(   8.)&$<$1.3          & 0.00&b\\
12048&15 41 23.73(0.05)& 66 20 35.8(0.3)&  5.7&   86.6&  103.3(   9.)&$<$1.3          & 0.00&b\\
00064&15 41 24.03(0.06)& 66 16 14.9(0.4)&  1.6&   66.2&  369.0(  48.)&  24.0          &22.96&i\\
12049&15 41 24.16(0.06)& 66 22 11.1(0.3)&  7.3&   67.6&   89.4(  10.)&$<$0.8          &22.03&i\\
00065&15 41 24.23(0.08)& 66 16 40.0(0.5)&  2.0&   40.3&   87.2(  19.)&   2.3x0.8p=95  & 0.00&b\\
00066&15 41 24.26(0.05)& 66 12 21.6(0.3)&  2.8&  105.8&  110.8(   8.)&$<$1.0          &22.87&i\\
25024&15 41 24.26(0.07)& 66 07 56.7(0.4)&  7.1&   42.7&   56.1(   9.)&$<$1.4          &21.30&i\\
25025&15 41 24.44(0.04)& 66 09 03.0(0.2)&  6.0&  135.2&  165.0(  10.)&$<$0.8          & 0.00&i\\
25026&15 41 24.91(0.05)& 66 05 25.6(0.3)&  9.6&   81.2&  130.6(  12.)&$<$1.4          &22.64&i\\
00067&15 41 25.49(0.07)& 66 14 41.8(0.4)&  1.2&   46.9&   47.3(   7.)&$<$1.4          & 0.00&i\\
12050&15 41 26.06(0.04)& 66 20 57.1(0.2)&  6.1&  139.9&  170.8(  10.)&$<$1.0          & 0.00&i\\
00068&15 41 26.16(0.07)& 66 13 41.4(0.4)&  1.8&   37.2&   38.0(   7.)&$<$1.4  & 0.00&b\\
12051&15 41 26.62(0.06)& 66 27 21.5(0.4)& 12.4&   75.8&  172.4(  21.)&   1.2x0p=5     &24.17&i\\
25027&15 41 26.82(0.07)& 66 02 54.1(0.4)& 12.2&   54.1&  143.9(  27.)&   1.8x0.2p=179 &23.75&i\\
00069&15 41 26.90(0.06)& 66 14 37.3(0.3)&  1.4&   82.0&   86.4(  10.)&   1.1x0p=8     & 0.00&b\\
25028&15 41 27.28(0.07)& 66 04 53.4(0.4)& 10.2&   43.9&   74.4(  12.)&$<$1.8          &23.35&i\\
00071&15 41 27.29(0.06)& 66 16 17.0(0.3)&  1.9&   63.4&   67.0(   8.)&$<$1.1          & 0.00&b\\
00072&15 41 27.37(0.06)& 66 17 40.9(0.4)&  3.0&   57.3&   83.4(  12.)&   1.3x0.6p=158 &23.02&i\\
00073&15 41 27.66(0.06)& 66 17 08.8(0.4)&  2.5&   55.6&   57.8(   7.)&$<$1.3          &24.74&i\\
00074&15 41 27.66(0.07)& 66 15 34.2(0.4)&  1.5&   37.7&   38.2(   7.)&$<$1.0          &25.32&i\\
25029&15 41 27.89(0.06)& 66 07 58.8(0.4)&  7.2&   54.6&   71.9(   9.)&$<$1.6          &23.27&i\\
00075&15 41 28.07(0.04)& 66 13 26.1(0.3)&  2.1&   89.0& 1318.0(  86.)&  18.0          &18.23&i\\
25030&15 41 28.12(0.07)& 66 05 16.6(0.4)&  9.8&   40.7&   66.6(  12.)&$<$1.9          & 2.90&i\\
12052&15 41 28.27(0.04)& 66 22 55.4(0.3)&  8.1&  107.1&  150.5(  11.)&$<$0.9          &23.56&i\\
00076&15 41 28.55(0.08)& 66 16 01.6(0.5)&  1.8&   41.9&   76.2(  17.)&   2.0x0.8p=15  &21.96&i\\
12053&15 41 28.78(0.02)& 66 22 02.7(0.1)&  7.2& 1004.4& 1323.3(  41.)&   0.4x0p=9     &24.26&i\\
12054&15 41 29.02(0.02)& 66 22 10.0(0.1)&  7.3&  744.1&  989.2(  31.)&   0.3x0p=9     & 0.00&b\\
25032&15 41 29.75(0.09)& 66 04 51.5(0.5)& 10.3&   38.8&  109.6(  32.)&   2.9x0p=94    &21.22&i\\
25033&15 41 29.78(0.08)& 66 10 02.8(0.5)&  5.2&   39.4&   71.4(  16.)&   1.6x0.7p=36  & 0.00&b\\
12108&15 41 30.62(0.07)& 66 25 39.0(0.4)& 10.8&   36.6&   65.4(  13.)&$<$1.7  &23.67&i\\
12055&15 41 31.12(0.06)& 66 22 55.4(0.3)&  8.1&   60.1&   84.8(  10.)&$<$1.2          & 0.00&i\\
25035&15 41 31.56(0.07)& 66 05 11.0(0.4)& 10.0&   38.5&   63.9(  12.)&$<$1.1  &23.76&c\\
12056&15 41 31.88(0.07)& 66 20 22.9(0.4)&  5.7&   50.3&   76.7(  15.)&   1.6x0p=67    &25.09&i\\
12057&15 41 33.21(0.07)& 66 19 53.1(0.4)&  5.3&   43.3&   50.5(   8.)&$<$1.0          & 0.00&i\\
00079&15 41 33.21(0.06)& 66 11 57.7(0.4)&  3.6&   49.1&   53.0(   8.)&$<$1.2          & 0.00&c\\
25036&15 41 33.29(0.07)& 66 08 33.3(0.4)&  6.7&   44.4&   56.8(   9.)&$<$1.5          &21.37&i\\
25037&15 41 33.42(0.06)& 66 09 36.2(0.4)&  5.7&   54.6&   65.6(   9.)&$<$1.3          &22.60&i\\
00080&15 41 33.58(0.07)& 66 13 31.3(0.4)&  2.5&   39.7&   41.2(   7.)&$<$1.7  & 0.00&c\\
00081&15 41 35.63(0.06)& 66 15 04.6(0.3)&  2.2&   74.1&  105.4(  13.)&   1.4x0.3p=118 & 0.00&i\\
00082&15 41 36.51(0.07)& 66 16 50.5(0.4)&  2.9&   42.5&   44.7(   7.)&$<$1.2          &25.07&i\\
25040&15 41 39.95(0.06)& 66 03 05.4(0.4)& 12.2&   52.8&  109.9(  15.)&$<$1.7          &20.36&i\\
12059&15 41 41.01(0.02)& 66 22 37.9(0.1)&  8.1&37676.0&57741.0(1732.)&  20.0          &20.67&i\\
00086&15 41 41.75(0.07)& 66 15 02.8(0.4)&  2.8&   54.1&   96.1(  17.)&   2.0          & 0.00&i\\
00087&15 41 41.92(0.04)& 66 11 10.3(0.2)&  4.8&  149.0&  170.0(   9.)&$<$0.6          &24.74&i\\
25041&15 41 42.46(0.06)& 66 06 22.4(0.4)&  9.1&   48.9&   75.4(  11.)&$<$1.4          &24.42&i\\
25042&15 41 42.79(0.05)& 66 06 46.8(0.3)&  8.7&   69.4&  103.7(  11.)&$<$1.5          &23.45&i\\
25043&15 41 42.93(0.06)& 66 05 58.2(0.4)&  9.5&   55.8&   89.2(  12.)&$<$1.8          &23.98&i\\
25044&15 41 43.34(0.07)& 66 04 33.5(0.4)& 10.9&   61.6&  140.4(  27.)&   2.3x0p=165   &23.24&i\\
12060&15 41 43.76(0.07)& 66 25 17.4(0.4)& 10.7&   52.7&  143.1(  26.)&   1.9x0.8p=27  &23.47&i\\
00090&15 41 44.84(0.07)& 66 13 48.3(0.4)&  3.3&   38.5&   41.3(   8.)&$<$1.7          &21.33&i\\
00091&15 41 45.03(0.05)& 66 15 11.3(0.3)&  3.1&   88.9&   94.7(   8.)&$<$1.1          &23.52&i\\
00092&15 41 45.29(0.08)& 66 13 57.0(0.5)&  3.3&   45.5&   52.7(  11.)&   1.4x0p=51    &24.32&i\\
00093&15 41 46.18(0.06)& 66 13 53.9(0.4)&  3.4&   57.6&   62.1(   8.)&$<$1.1          &22.39&i\\
00094&15 41 46.24(0.07)& 66 11 15.7(0.4)&  5.0&   39.1&   45.1(   8.)&$<$1.2          &20.48&i\\
00095&15 41 47.26(0.07)& 66 17 08.6(0.4)&  4.0&   39.3&   43.2(   8.)&$<$1.6          &20.27&i\\
00096&15 41 47.36(0.09)& 66 11 25.4(0.5)&  4.9&   42.5&  140.0(  37.)&   9.0          & 0.00&i\\
00097&15 41 48.80(0.03)& 66 11 36.8(0.2)&  4.9&  330.4& 2112.0(  83.)&  37.0          & 0.00&i\\
12062&15 41 49.65(0.06)& 66 25 54.2(0.4)& 11.5&   58.5&  112.4(  14.)&$<$1.5          &22.47&i\\
25045&15 41 50.24(0.07)& 66 05 32.4(0.4)& 10.1&   45.7&   77.6(  12.)&$<$1.4          &24.86&i\\
00100&15 41 51.57(0.06)& 66 16 43.1(0.4)&  4.2&   49.4&   55.0(   8.)&$<$1.1          &25.14&i\\
12063&15 41 52.56(0.06)& 66 27 25.9(0.4)& 13.0&   48.6&  110.3(  16.)&$<$1.8          &21.32&i\\
25046&15 41 52.73(0.06)& 66 05 29.3(0.4)& 10.3&   52.3&   90.2(  12.)&$<$1.3          &24.20&i\\
12064&15 41 53.72(0.04)& 66 20 30.7(0.2)&  6.8&  244.7& 1132.3(  59.)&  18.0       & 0.00&c\\
12066&15 41 54.39(0.03)& 66 23 23.0(0.2)&  9.3&  316.4&  496.4(  19.)&$<$0.5          & 0.00&c\\
00102&15 41 54.56(0.05)& 66 16 24.0(0.3)&  4.3&   69.9&   78.5(   8.)&$<$1.2    &22.61&i\\
00103&15 41 54.73(0.09)& 66 15 52.9(0.5)&  4.2&   39.2&   75.6(  22.)&   3.0x1.1p=76  &23.90&i\\
19001&15 41 56.96(0.03)& 66 12 10.9(0.2)&  5.2&  320.4&  376.7(  14.)&$<$0.7          &23.01&i\\
19002&15 41 57.34(0.06)& 66 13 22.2(0.4)&  4.7&   54.3&   62.2(   8.)&$<$0.8          & 0.00&c\\
19003&15 41 57.43(0.09)& 66 12 56.3(0.5)&  4.8&   37.6&   49.2(  13.)&   1.5x0p=75    &19.95&i\\
19005&15 41 58.27(0.07)& 66 13 11.9(0.4)&  4.8&   45.7&   52.8(   8.)&$<$1.0          &25.84&i\\
13001&15 41 59.29(0.06)& 66 23 26.3(0.3)&  9.6&   61.1&   98.5(  12.)&$<$1.3          &23.47&i\\
19006&15 42 00.00(0.07)& 66 18 43.0(0.4)&  5.9&   50.5&   96.4(  16.)&   1.5x0.8p=119 &22.82&i\\
19007&15 42 00.60(0.07)& 66 18 09.3(0.4)&  5.6&   43.8&   52.5(   9.)&$<$1.4  & 0.00&b\\
26004&15 42 01.11(0.08)& 66 10 11.4(0.5)&  6.8&   38.3&   58.8(  15.)&   1.6x0p=67    &20.21&i\\
19009&15 42 01.41(0.05)& 66 14 31.1(0.3)&  4.8&   71.9&   83.1(   8.)&$<$1.2          & 0.00&i\\
19010&15 42 02.18(0.02)& 66 13 00.5(0.1)&  5.2& 5618.7&13249.0( 399.)&   8.6       &25.00&i\\
26005&15 42 02.48(0.07)& 66 06 46.8(0.4)&  9.6&   47.2&   76.6(  12.)&$<$2.0          &22.09&i\\
13002&15 42 03.07(0.07)& 66 20 15.2(0.4)&  7.2&   47.0&   62.8(  10.)&$<$1.7       & 0.00&c\\
19012&15 42 03.38(0.02)& 66 17 40.2(0.1)&  5.6& 2320.5& 2804.0(  85.)&$<$0.2          &26.02&i\\
19013&15 42 03.45(0.05)& 66 15 53.3(0.3)&  5.1&  102.4&  119.9(   9.)&$<$1.1          & 0.00&c\\
19014&15 42 05.59(0.05)& 66 15 27.7(0.3)&  5.2&  102.9&  121.8(   9.)&$<$1.1          & 0.00&c\\
19015&15 42 05.86(0.07)& 66 13 30.9(0.4)&  5.4&   35.4&   41.9(   8.)&$<$1.3          &23.71&i\\
19016&15 42 05.88(0.06)& 66 16 52.7(0.3)&  5.5&   63.0&   75.9(   9.)&$<$1.3          &22.55&i\\
13106&15 42 06.70(0.09)& 66 19 11.6(0.5)&  6.8&   37.1&  107.1(  31.)&   3.9x2.0p=102 &22.14&i\\
19019&15 42 07.53(0.02)& 66 19 00.6(0.1)&  6.7& 2442.4& 3562.1( 108.)&   1.0x0p=136   &24.18&i\\
19020&15 42 08.22(0.02)& 66 18 42.9(0.1)&  6.6& 1246.6& 1602.9(  49.)&$<$0.4          &24.94&i\\
26007&15 42 08.74(0.05)& 66 06 02.1(0.3)& 10.5&   72.4&  128.8(  13.)&$<$1.5          &23.72&i\\
19021&15 42 09.19(0.05)& 66 13 49.4(0.3)&  5.7&   82.4&  105.3(  11.)&   0.9x0p=161   &26.07&i\\
19022&15 42 09.77(0.09)& 66 14 51.4(0.5)&  5.6&   47.8&   81.4(  22.)&   3.4x0p=45    &20.65&i\\
19023&15 42 10.11(0.06)& 66 12 17.1(0.4)&  6.3&   47.8&   60.5(   9.)&$<$1.4          &23.62&i\\
13005&15 42 10.72(0.05)& 66 21 13.8(0.3)&  8.4&  113.6&  174.6(  14.)&   1.0x0p=172   & 0.00&b\\
19024&15 42 10.87(0.08)& 66 11 09.9(0.5)&  6.9&   45.8&   94.8(  22.)&   2.4 lrms=6.8 & 0.00&b\\
13006&15 42 11.08(0.05)& 66 25 13.9(0.3)& 11.7&  135.2&  327.8(  24.)&   1.4x0.4p=43  &24.96&i\\
19025&15 42 11.09(0.07)& 66 12 07.8(0.4)&  6.4&   40.3&   50.9(   9.)&$<$1.5          & 0.00&b\\
26008&15 42 11.39(0.06)& 66 04 36.5(0.3)& 11.9&   66.0&  135.0(  15.)&$<$1.6          &20.33&i\\
13007&15 42 11.54(0.06)& 66 25 43.6(0.3)& 12.2&   62.0&  129.2(  15.)&$<$2.1          &25.72&i\\
26009&15 42 11.87(0.07)& 66 05 41.7(0.4)& 11.0&   39.9&   74.4(  13.)&$<$1.1          &25.56&i\\
19026&15 42 13.03(0.08)& 66 13 26.5(0.5)&  6.2&   34.3&   43.1(   9.)&$<$1.8          & 0.00&i\\
19027&15 42 14.76(0.06)& 66 17 07.7(0.3)&  6.5&   80.1&  105.2(  11.)&   0.9x0p=53    &20.78&i\\
13010&15 42 14.82(0.03)& 66 22 49.6(0.2)&  9.9&  271.8&  455.1(  18.)&$<$0.7          & 0.00&i\\
26011&15 42 15.09(0.05)& 66 10 17.3(0.3)&  7.8&   73.6&  103.7(  10.)&$<$1.0          & 0.00&b\\
19028&15 42 15.37(0.07)& 66 18 09.7(0.4)&  6.9&   47.2&  104.2(  18.)&   1.5x1.2p=165 & 0.00&b\\
19029&15 42 15.39(0.05)& 66 17 53.1(0.3)&  6.8&   82.7&  108.5(  10.)&$<$1.1          &23.41&i\\
19030&15 42 15.54(0.07)& 66 16 41.5(0.4)&  6.4&   47.1&   60.3(   9.)&$<$1.0          &23.49&i\\
13035&15 42 15.94(0.07)& 66 19 52.7(0.4)&  7.9&   44.0&   62.3(  10.)&$<$1.3          & 0.00&c\\
19031&15 42 16.00(0.08)& 66 14 44.1(0.5)&  6.2&   45.8&   70.3(  14.)&   1.4x0p=20    &23.36&i\\
13011&15 42 16.40(0.02)& 66 19 33.0(0.1)&  7.7& 3558.2&23352.0( 703.)&  15.0          &22.27&i\\
19032&15 42 16.91(0.07)& 66 15 28.2(0.4)&  6.3&   47.1&   60.1(   9.)&$<$1.3          & 0.00&i\\
19033&15 42 17.51(0.06)& 66 12 02.7(0.3)&  7.1&   67.7&   90.8(  10.)&$<$1.0          &23.16&i\\
26012&15 42 17.93(0.05)& 66 02 51.2(0.3)& 13.8&   80.6&  203.2(  19.)&$<$1.9          & 0.00&c\\
19034&15 42 18.43(0.06)& 66 18 04.2(0.4)&  7.2&   51.0&   68.8(  10.)&$<$0.9          & 0.00&c\\
13012&15 42 21.89(0.07)& 66 21 08.6(0.4)&  9.2&   46.5&   73.3(  11.)&$<$1.9          &20.14&i\\
26014&15 42 21.68(0.04)& 66 07 19.1(0.3)& 10.3&  277.4&  589.3(  41.)&  12.0          &25.55&i\\
13013&15 42 22.89(0.07)& 66 22 24.2(0.4)& 10.1&   39.6&   67.8(  12.)&$<$1.9  & 0.00&b\\
26015&15 42 24.18(0.05)& 66 10 05.5(0.3)&  8.6&   74.9&  113.7(  11.)&$<$1.2          & 0.00&i\\
26016&15 42 24.69(0.04)& 66 10 12.3(0.2)&  8.6&  136.4&  206.7(  12.)&$<$0.9          &19.68&i\\
26017&15 42 25.00(0.04)& 66 09 17.3(0.2)&  9.2&  151.1&  240.2(  13.)&$<$0.9          &21.32&i\\
19036&15 42 25.03(0.07)& 66 13 06.3(0.4)&  7.4&   49.5&   78.6(  14.)&   1.1x0p=38    & 0.00&c\\
19037&15 42 26.15(0.03)& 66 15 39.1(0.2)&  7.3&  372.5&  510.3(  18.)&$<$0.4          & 0.00&i\\
13018&15 42 28.26(0.06)& 66 22 02.8(0.3)& 10.2&   65.7&  114.4(  13.)&$<$1.0          & 0.00&b\\
26018&15 42 28.89(0.06)& 66 02 56.4(0.4)& 14.3&   51.4&  138.1(  19.)&$<$2.1          &25.43&i\\
19038&15 42 29.73(0.04)& 66 14 27.3(0.2)&  7.6&  171.7&  242.3(  12.)&$<$0.5          &21.45&i\\
19040&15 42 31.56(0.08)& 66 14 26.8(0.5)&  7.8&   37.6&  244.1(  54.)&   5.1x2.6p=156 &18.57&i\\
19041&15 42 32.13(0.07)& 66 16 19.6(0.4)&  8.0&   37.9&   54.8(  10.)&$<$1.7          &22.18&i\\
19042&15 42 36.61(0.05)& 66 11 04.7(0.3)&  9.2&   72.8&  117.3(  12.)&$<$1.1          & 0.00&c\\
26020&15 42 36.94(0.07)& 66 05 13.3(0.4)& 12.9&   42.8&   98.7(  16.)&$<$1.3          & 0.00&b\\
26021&15 42 37.26(0.06)& 66 02 44.8(0.4)& 14.9&   49.7&  144.9(  21.)&$<$1.3          &21.26&i\\
19044&15 42 39.13(0.06)& 66 13 26.2(0.4)&  8.7&   53.5&   82.7(  11.)&$<$1.1          &19.51&i\\
19045&15 42 39.79(0.03)& 66 18 35.9(0.2)&  9.3&  224.4&  364.5(  16.)&$<$0.7          & 0.00&b\\
26022&15 42 40.06(0.06)& 66 07 03.1(0.4)& 11.8&   51.7&  106.4(  15.)&$<$1.6          & 0.00&c\\
13021&15 42 40.59(0.07)& 66 26 40.0(0.4)& 14.5&   38.4&  108.1(  20.)&$<$1.4  &24.72&i\\
13022&15 42 41.42(0.06)& 66 24 33.7(0.4)& 13.0&   70.6&  334.4(  47.)&   2.6x0.8p=111 &19.51&i\\
13122&15 42 41.09(0.05)& 66 24 36.7(0.3)& 13.0&  101.0&  404.7(  36.)&   1.7x1.4p=127 &18.75&i\\
13036&15 42 42.22(0.08)& 66 20 36.7(0.5)& 10.5&   40.4&  132.1(  31.)&   2.2x0.8p=162 &20.20&i\\
19046&15 42 43.15(0.08)& 66 15 05.8(0.5)&  9.0&   51.5&  209.5(  45.)&   5.1x1.8p=6   &22.90&i\\
26023&15 42 46.78(0.06)& 66 08 41.4(0.3)& 11.3&   65.2&  128.5(  14.)&$<$1.2          &24.64&i\\
19049&15 42 47.61(0.03)& 66 14 22.6(0.2)&  9.4&  417.3&  690.8(  24.)&$<$0.6          &22.06&i\\
13024&15 42 49.46(0.07)& 66 23 22.8(0.4)& 12.7&   41.8&   94.8(  16.)&$<$1.5  & 0.00&b\\
19050&15 42 51.21(0.07)& 66 12 26.9(0.4)& 10.1&   44.1&   77.9(  13.)&$<$1.4          & 0.00&c\\
13026&15 42 51.61(0.04)& 66 24 13.2(0.3)& 13.4&  114.4&  281.2(  19.)&$<$1.1          & 0.00&i\\
13027&15 42 51.79(0.08)& 66 20 37.3(0.5)& 11.3&   42.5&  146.2(  33.)&   2.2x1.2p=69  &24.64&i\\
19051&15 42 52.50(0.07)& 66 19 02.5(0.4)& 10.7&   43.1&   77.6(  13.)&$<$1.8          & 0.00&i\\
19052&15 42 53.18(0.06)& 66 16 58.5(0.4)& 10.2&   56.0&   99.0(  13.)&$<$1.1          &24.66&i\\
26025&15 42 54.38(0.04)& 66 03 35.2(0.2)& 15.3&  226.5&  801.1(  39.)&   1.1x0.5p=144 &23.33&i\\
19053&15 42 54.38(0.06)& 66 18 08.1(0.4)& 10.6&   54.4&   99.8(  13.)&$<$1.5          &20.52&i\\
19054&15 42 54.64(0.07)& 66 11 15.1(0.4)& 10.8&   38.1&   69.4(  13.)&$<$2.3          &20.52&i\\
19055&15 42 55.98(0.07)& 66 12 12.6(0.4)& 10.7&   62.2&  148.3(  22.)&   1.5x0.5p=110 &24.19&i\\
19056&15 42 58.49(0.08)& 66 11 46.7(0.5)& 11.0&   37.8&  103.9(  26.)&   1.8x0.8p=118 &23.29&i\\
13028&15 42 59.97(0.03)& 66 24 24.7(0.2)& 14.2&  277.6&  795.2(  33.)&   0.8x0p=68    &22.58&i\\
19058&15 43 01.07(0.03)& 66 15 16.9(0.2)& 10.8&  386.0&  729.3(  26.)&$<$0.5          &20.37&i\\
19060&15 43 05.79(0.07)& 66 17 20.4(0.4)& 11.5&   42.7&   83.1(  14.)&$<$1.7  &26.85&i\\
26026&15 43 08.89(0.03)& 66 10 43.6(0.2)& 12.4&  382.0&  852.2(  30.)&$<$0.8          &21.98&i\\
13031&15 43 09.31(0.06)& 66 27 07.1(0.3)& 16.7&   65.3&  245.0(  27.)&$<$1.9          &23.42&i\\
26027&15 43 09.43(0.08)& 66 05 09.2(0.5)& 15.3&   34.8&  108.8(  22.)&$<$2.0          &24.24&i\\
13032&15 43 09.84(0.02)& 66 20 06.4(0.1)& 12.7& 1067.3& 2458.7(  76.)&$<$0.4          &24.10&i\\
19063&15 43 10.49(0.04)& 66 17 57.0(0.2)& 12.1&  141.3&  305.4(  18.)&$<$1.2          &20.28&i\\
13033&15 43 11.44(0.04)& 66 20 57.7(0.2)& 13.2&  187.2&  509.3(  31.)&   1.5x0p=58    &22.71&i\\
19064&15 43 12.16(0.07)& 66 11 29.0(0.4)& 12.4&   42.4&   95.5(  16.)&$<$1.3          &25.14&i\\
26028&15 43 12.75(0.05)& 66 04 19.4(0.3)& 16.1&   74.1&  257.6(  26.)&$<$2.5          &23.92&i\\
19065&15 43 13.41(0.05)& 66 16 58.3(0.3)& 12.2&  103.7&  227.0(  17.)&$<$1.0          &24.70&i\\
26029&15 43 13.71(0.06)& 66 07 58.9(0.4)& 14.0&   67.5&  229.0(  31.)&   1.4x0.7p=109 &19.09&i\\
26030&15 43 14.01(0.05)& 66 08 53.4(0.3)& 13.6&   82.7&  211.7(  19.)&$<$1.7          &24.96&i\\
19067&15 43 14.63(0.07)& 66 15 15.2(0.4)& 12.1&   35.1&   76.9(  16.)&$<$2.3  &23.68&i\\
\enddata
\end{deluxetable}

\end{document}